# Contextual Minimum-Norm Estimates (CMNE): A Deep Learning Method for Source Estimation in Neuronal Networks

**Christoph Dinh[1,2*], John GW Samuelsson[1*], Alexander Hunold[3], Matti S Hämäläinen[1], Sheraz Khan[1]**

[1]*Massachusetts General Hospital - Massachusetts Institute of Technology - Harvard Medical School; Harvard-MIT Division of Health Sciences and Technology, Athinoula A. Martinos Center for Biomedical Imaging, 149 13th St., Charlestown, MA 02129, USA*
[2]*Institute for Medical Engineering, Research Campus STIMULATE, Otto-von-Guericke University, Magdeburg, GER*
[3]*Institute for Biomedical Engineering and Informatics, Technische Universität Ilmenau, 98693 Ilmenau, GER*

**Abstract:** Magnetoencephalography (MEG) and Electroencephalography (EEG) source estimates have thus far mostly been derived sample by sample, i.e., independent of each other in time. However, neuronal assemblies are heavily interconnected, constraining the temporal evolution of neural activity in space as detected by MEG and EEG. The observed neural currents are thus highly context dependent. Here, a new method is presented which integrates predictive deep learning networks with the Minimum-Norm Estimates (MNE) approach. Specifically, we employ Long Short-Term Memory (LSTM) networks, a type of recurrent neural network, for predicting brain activity. Because we use past activity (context) in the estimation, we call our method Contextual MNE (CMNE). We demonstrate that these contextual algorithms can be used for predicting activity based on previous brain states and when used in conjunction with MNE, they lead to more accurate source estimation. To evaluate the performance of CMNE, it was tested on simulated and experimental data from human auditory evoked response experiments.

*Keywords: MEG, EEG, Distributed Source Estimation, grid-based Markov localization, dSPM, LSTM, Deep Learning*

## 1 INTRODUCTION

Neural currents in the brain can be estimated from MEG/EEG recordings by solving the inverse problem (Hamalainen et al. 1993; Mosher, Leahy, and Lewis 1999). The inverse problem is ill-posed: several current distributions can produce the same or similar electric and magnetic fields outside the head and the estimates therefore become sensitive to measurement noise (Hamalainen et al. 1993; Helmholtz 1853). These difficulties limit the spatial resolution and reliability of neural current estimates derived from MEG/EEG signals. To deal with this ill-posedness of the inverse problem, constraints limiting the space of possible neural current configurations and regularization are often used. Solving the inverse problem requires a forward model that calculates the MEG/EEG signals from given current distributions in the brain (Sarvas 1987; Mosher, Leahy, and Lewis 1999; Stenroos, Hunold, and Haueisen 2014).

*\*Both authors contributed equally to this work*
E-mail: christoph.dinh@mne-cpp.org (corresponding author); johnsam@mit.edu; alexander.hunold@tu-ilmenau.de; msh@nmr.mgh.harvard.edu; sheraz@nmr.mgh.harvard.edu

Popular methods for solving the inverse problem include discrete current dipole models (Schneider 1972; Scherg and Cramon 1985; Mosher, Lewis, and Leahy 1992; Leahy et al. 1998) as well as distributed current models (Hamalainen and Ilmoniemi 1994; Uutela, Hamalainen, and Somersalo 1999; Baillet, Mosher, and Leahy 2001; Stenbacka et al. 2002). Importantly, most source estimation methods are derived sample by sample, i.e., without assuming any relationship between the current distributions across time.

Ways of dealing with the ill-posedness of the inverse problem include the introduction of penalty terms (regularization) for current sources with large power or amplitude (Hamalainen and Ilmoniemi 1994; Uutela, Hamalainen, and Somersalo 1999), assuming spatial smoothness (Pascual-Marqui, Michel, and Lehmann 1994; Dinh et al. 2015, 2017) and imposing focal estimates (Gorodnitsky, George, and Rao 1995). Spatially sparse source reconstructions have also been attained by using Bayesian methods that employ the current source covariance (Phillips et al. 2005; Mattout et al. 2006) or by computing posterior distributions based on hierarchical priors (Sato et al. 2004; Nummenmaa et al. 2007; Costa et al. 2017). These methods, however, commonly assume that the prior probability distribution of the sources is static. This assumption ignores the temporal structure of the underlying neural activity (Buzsáki and Draguhn 2004) that could be used to mitigate the ill-posedness of the inverse problem.

More recent Bayesian methods for source estimation incorporate temporal smoothness constraints, which specify various prior distributions for the dipole sources in space and time (Baillet and Garnero 1997; Greensite 2003; Trujillo-Barreto, Aubert-Vázquez, and Penny 2008; Zumer et al. 2008; Friston et al. 2008; Bolstad, Veen, and Nowak 2009; Ou, Hamalainen, and Golland 2009; Limpiti et al. 2009). Mixed-Norm estimates have been introduced imposing stationarity of the source estimates within a time window and quasinorm penalties to promote spatial sparsity (Strohmeier et al. 2016). Estimation methods employing linear state-space models either apply spatially-independent approximations (Galka et al. 2004; Yamashita et al. 2004) or a priori fix model specific parameters (Long et al. 2011) to reduce their computational burden. These methods incorporate temporal structure by including spatiotemporal dynamic constraints, such as space-time separability or spatial independence.

Recent developments introduce source estimation methods based on more realistic spatiotemporal dynamic models using Kalman filters (Kalman 1960), which take local cortical interactions considering neuroanatomical and electrophysiological conditions into account (Lamus et al. 2012; Pirondini et al. 2017). Furthermore, methods using structured sparse priors defined in the time-frequency domain have been proposed. This approach is able to accommodate the non-stationary and transient nature of brain signals (Gramfort et al. 2013). In summary, the above models incorporate temporal structure while assuming constrained spatiotemporal interactions, e.g., space-time separability or spatial independence. Although these models have shown a lot of promise, the potential of incorporating long range dynamic interactions, i.e., the overall *context* (past) of the activity, to improve inverse solutions has remained unexplored.

Several studies have shown that dynamic spatiotemporal interactions are a central feature of brain activity. Intracranial recordings show strong local spatial correlations up to distances of 10 mm along the cortical surface (Bullock et al. 1995; Destexhe, Contreras, and Steriade 1999; Leopold, Murayama, and Logothetis 2003). Physiologically motivated spatiotemporal models of neuronal networks have been able to predict aspects of EEG and MEG recordings (Gross et al. 2001; Jirsa et al. 2002; Wright et al. 2003; Robinson et al. 2005; Izhikevich and Edelman 2008). Temporally coherent fluctuations in activation in widely distributed cortical networks have also been shown in fMRI studies (Fox et al. 2005; Fox and Raichle 2007; Gusnard and Raichle 2001; Raichle et al. 2001). These observations indicate that cortical activity depend on past brain activity, i.e., their context.

Recent advances in machine learning have focused on sequential data sets, e.g., reinforcement learning, enabling recognition of the context of the data (LeCun, Bengio, and Hinton 2015; Schmidhuber 2015). These new contextual capabilities have been demonstrated to significantly improve classification accuracy in human-machine interaction, e.g., text and speech recognition (Graves, Mohamed, and Hinton 2013; Cho et al. 2014; Bahdanau, Cho, and Bengio 2015). We hypothesize that these contextual techniques can be used to model brain activity (Wu, Nagarajan, and Chen 2016) and, consequently, utilized to predict neural activity.

In this article, we present a new grid-based Markov localization method, called Contextual Minimum Norm Estimates (CMNE), which integrates dynamical statistic parameter mapping (dSPM), a well-established distributed source estimation method, with long short-term memory (LSTM) neural networks (Hochreiter and Schmidhuber 1997) trained on a larger block of MEG/EEG data, thus incorporating contextual information. To test the efficacy of this method, we try it on simulated and real human MEG/EEG data from auditory steady state response experiments and compare it to other commonly used methods.

## 2 METHODS

### 2.1 Contextual Minimum-Norm Estimates

MEG/EEG signals are commonly linked to their source generators via a forward model $G$;

$$y_t = Gq_t + n_t, \qquad (1)$$

where the vector $y_t$ represents the signal in the senor array at time $t$, $G$ is the time-invariant gain matrix, the vector $q_t$ contains the amplitudes of current dipole sources, and $n_t$ is the noise. In our model, freely oriented current dipoles were evenly distributed over the white matter surface.

In the following we will use the whitened measured signals $\tilde{y}_t$ and gain matrix $\tilde{G}$:

$$\tilde{y}_t = C_n^{-1/2} y_t \text{ and } \tilde{G} = C_n^{-1/2} G, \qquad (2)$$

where $C_n$ is the estimated noise covariance matrix (Engemann and Gramfort 2015). Many source estimates can be expressed as the solution of a minimization problem of the form:

$$\hat{q}_t = \arg \min_{q_t} \left( \left\| \tilde{y}_t - \tilde{G} q_t \right\| + f(q_t) \right), \qquad (3)$$

where the first term is the norm of the differences between the measured signal $\tilde{y}_t$ and the predicted signal $\tilde{G} q_t$ based on the model $\tilde{G}$ and current source configuration $q_t$, while $f(q)$ incorporates a priori assumptions or regularization. A common approach for solving this inverse problem is the minimum-norm estimate (MNE) (Hamalainen and Ilmoniemi 1994). In MNE, the $\ell^2$-norm is used in the error term and $f(q_t) = \lambda^2 q_t^T C_R^{-1} q_t$, where $C_R$ is the (assumed) source covariance matrix and $\lambda$ is a hyperparameter adjusting the relative weight of the two terms in Eq. (3). The solution to this minimization problem can be written as a

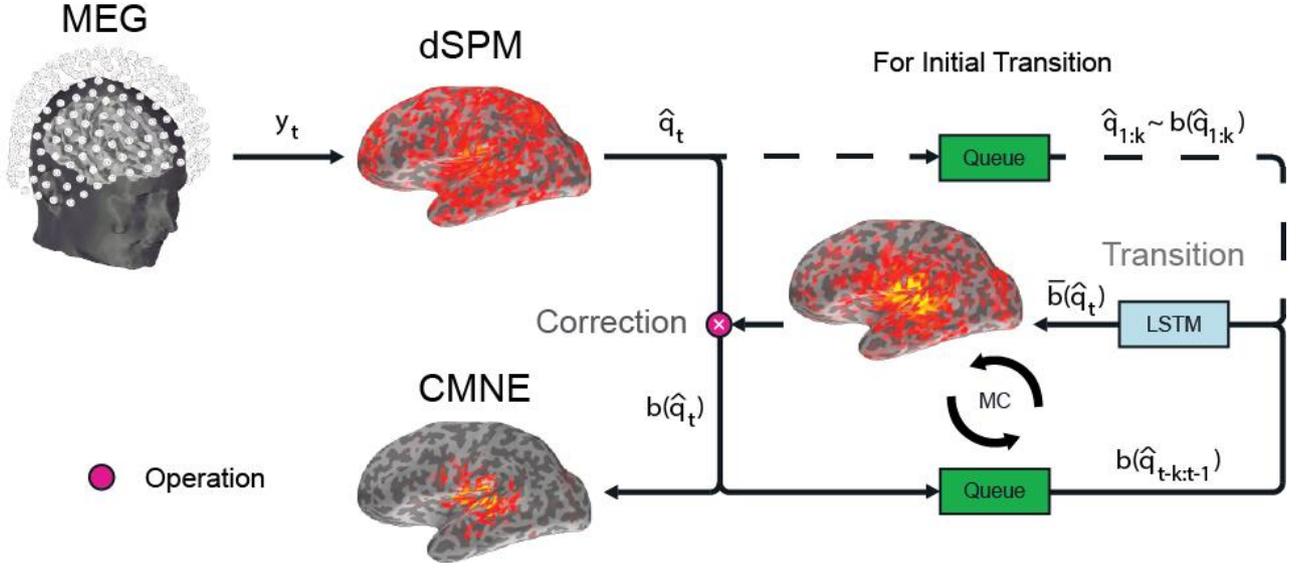

*Figure 1: In the first step, dSPM is employed to estimate the source space activation $\hat{q}_t$ of the sensor space measurement $y_t$. The dSPM activation distribution $\hat{q}_t$ is then corrected by the predicted belief distribution $\bar{b}(\hat{q}_t)$, yielding the posterior belief distribution $b(\hat{q}_t)$. The prior posterior belief distributions $b(\hat{q}_{t-k:t-1})$ are also the input for the transition function $u_t$ which calculates the prediction $\bar{b}(\hat{q}_t)$ using an LSTM network. This forms an iterative circle, i.e., a Markov chain (MC) (Dinh 2015). In the calculation of the first k belief distributions $b(\hat{q}_{1:k})$, no correction is applied. In other words, $b(\hat{q}_{1:k})$ equals the dSPM source estimates $\hat{q}_{1:k}$, since the transition model is based on an LSTM network that require k samples to calculate an initial prediction. A memory queue stores the previous k samples.*

product of the measured data $\tilde{y}_t$ and a matrix $M$ that we call the linear inverse operator:

$$\hat{q}_t = \arg\min_{q_t} \left( (\tilde{y}_t - \tilde{G}q_t)^T (\tilde{y}_t - \tilde{G}q_t) + \lambda^2 q_t^T C_R^{-1} q_t \right) = M\tilde{y}_t, \quad (4)$$

where

$$M = C_R \tilde{G}^T (\tilde{G} C_R \tilde{G}^T + \lambda^2 I)^{-1}. \quad (5)$$

One way to set the regularization parameter $\lambda$ is to relate it to the SNR of the whitened data. This approach is described in detail by (Lin et al. 2006). The kernel $M$ is applied sample-wise for each measurement $\tilde{y}_t$, computing source estimates $\hat{q}_t$ independent of each other in time.

In our Contextual MNE (CMNE) method, the source covariance matrix $C_R$ is assumed to be time-dependent, $C_R = C_R(t)$, and obtained from a variant of a discrete Bayes filter called grid-based Markov localization, making the current source estimate $\hat{q}_t$ dependent on previous source estimates $\{\hat{q}_{t-k}\}_{k=1,..,n}$ for some $n \geq 1$. It combines two models; a measurement model and a transition model, which will be explained in detail in the next section.

In the following, we will use dynamic statistical parametric mapping (dSPM) (Dale et al. 2000) for initial, sample-wise, source estimation $\hat{q}_t$. dSPM is a noise-normalized variant of MNE. We then use a transition model estimating the activation change $u_t$ based on previous source estimations $\{\hat{q}_{t-k}\}_{k=1,..,n}$, i.e., contextual information, to correct this source estimate $\hat{q}_t$, giving us the CMNE solution $b(\hat{q}_t)$. The transition model is based on LSTM networks and will be explained in greater detail in section 2.3.

## 2.2 Grid-based Markov localization

Here source activity is modeled as a nonparametric probability distribution over a cortical grid, i.e., a histogram distribution where the activity of each source point in the grid over the cortical mantle is represented by a single probability value. This model allows for solving the inverse problem using a grid-based Markov localization approach. It is an example of discrete Bayes filters applied to the localization problem (Burgard, Fox, and Thrun 1999). The grid-based Markov localization approach incorporates a source estimation method and a transition model. dSPM applied sample-wise on each measurement point $y_t$ independently is the source estimation method. A source activity *prediction* based on the past CMNE source estimates form the transition model. The two models are connected in a Markov chain. The predicted source activity distribution is called *predicted belief distribution*, denoted by $\bar{b}(\hat{q}_t)$, and is corrected by the dSPM estimate based on the current measurement $y_t$. This corrected source distribution is called the *posterior belief distribution*, denoted $b(\hat{q}_t)$, and is the distribution we use as source activity estimation;

$$b(\hat{q}_t) = p(\hat{q}_t | y_{1:t}, u_{2:t}), \quad (6)$$

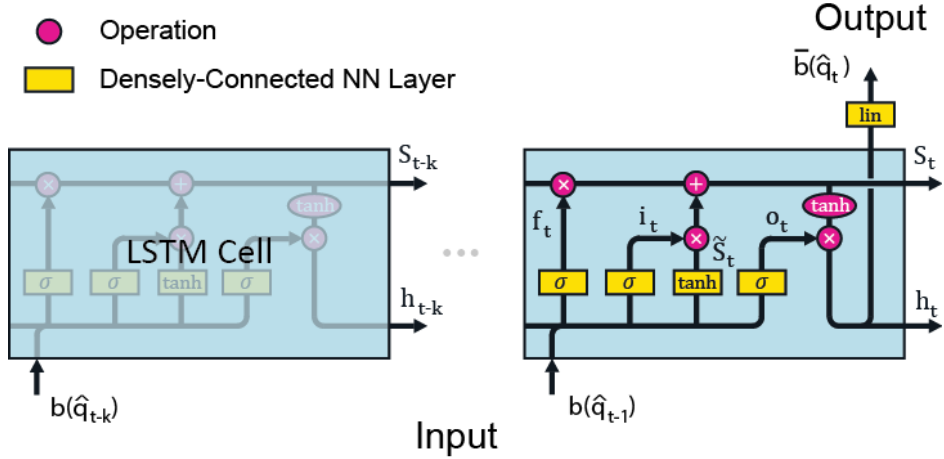

*Figure 2: The LSTM architecture with a subsequent regular densely-connected neuronal network (NN) layer containing a linear activation which estimates the belief distribution of the LSTM states. One LSTM cell consists of several densely-connected NN layers: (i) $f_t$, the "forget gate layer", is a NN with a sigmoid activation function. This layer gates the previous cell state $0 \leq S_{t-1} \leq 1$ depending on the input of the previous cell output $h_{t-1}$ and the previous belief $b(\hat{q}_{t-1})$. (ii) $i_t$, the "input gate layer", is also a sigmoid layer with the same inputs as $f_t$. It determines which cell state values to update. (iii) $\tilde{S}_t$ is the subsequent densely-connected NN layer with a tanh activation function, which creates candidate cell state values gated by $i_t$. The state $S_t$ of the current cell is formed by "forgetting" outdated information of the previous cell state $S_{t-1}$ through a multiplication with $f_t$ followed by an update with the gated new cell state candidates $i_t\tilde{S}_t$. The actual output of the LSTM cell $\bar{b}(\hat{q}_t)$ is a filtered subset of the cell state $S_t$. (iv) $o_t$ is a densely-connected NN layer activated by a sigmoid function. This layer decides which cell states to output. It gates the tanh scaled cell state $S_t$. See also (Olah 2015)*

where $\hat{q}_t$ is a column vector of length $\rho$, the number of grid points in the cortical source space; $y_t$ is a column vector of length $m$, the number of measurement channels; and $u_{2:t}$ is the set of all past activation changes.

Figure 1 shows an overview of this approach. Note that we use the convention that $u_t$ is the activation change from state $t-1$ to state $t$. $b(\hat{q}_t)$ is computed by incorporating the measurement $y_t$ based on the Dempster–Shafer theory (Dempster 1967; Shafer 1976; Fine 1977). The belief $b(\hat{q}_t)$ is a vector with the same dimensions as the source estimation $\hat{q}_t$.

The predicted belief distribution $\bar{b}(\hat{q}_t)$ before incorporating the current measurement $y_t$, right after the transition $u_t$, is the probability of source estimation $\hat{q}_t$ given all prior measurements $y_{1:t-1}$ and all activation changes $u_{2:t}$;

$$\bar{b}(\hat{q}_t) = p(\hat{q}_t | y_{1:t-1}, u_{2:t}). \quad (7)$$

This predicted belief distribution is also simply called *prediction* and has the same dimensionality as the belief $\bar{b}(\hat{q}_t)$. The predicted belief distribution is given by a trained long short-term memory (LSTM) network (Hochreiter and Schmidhuber 1997; Gers, Schmidhuber, and Cummins 2000), see Sec.2.3.

In accordance with the Dempster–Shafer theory (Dempster 1967; Shafer 1976; Fine 1977), the calculation of $b(\hat{q}_t)$ from $\bar{b}(\hat{q}_t)$, also called *correction*, is:

$$b(\hat{q}_t) = \eta p(\hat{q}_t | y_t) \bar{b}(\hat{q}_t), \quad (8)$$

where $\eta$ is a scalar that normalizes the posterior probability distribution, which employs $\bar{b}(\hat{q}_t)$ as the prior. The combination of prediction and measurement is forming a Markov Chain (MC) as illustrated in Figure 1.

## 2.3 LSTM

Long Short-Term Memory (LSTM) networks (Hochreiter and Schmidhuber 1997) are a special kind of recurrent neural network (RNN), capable of learning long-term dependencies. LSTMs are known to perform well in sequential processing tasks in which hierarchical decompositions, i.e., subprocesses, may exist but are not known in advance (Gers, Schmidhuber, and Cummins 2000), e.g., language discrimination with unclear linguistic theory of hierarchical structure (Cummins, Gers, and Schmidhuber 1999). LSTM memory blocks and their forget gates can develop into internal oscillators or timers, allowing the recognition and generation of hierarchical rhythmic patterns (Gers, Schmidhuber, and Cummins 2000). These desirable properties led to the idea of applying LSTM networks to the source estimation problem in M/EEG. Here LSTM networks are used to predict the source activation distribution $\bar{b}(\hat{q}_t)$ based on the past estimates $b(\hat{q}_{t-k:t-1})$.

Our transition model is formed by an LSTM network followed by a linear transformation mapping the output of the last LSTM cell $h_t$ to the predicted belief distribution $\bar{b}(\hat{q}_t)$ (Figure 2). The LSTM network itself consists of an LSTM cell sequence, each cell $i$ having a source activation distribution of the corresponding past time step $b(\hat{q}_{t-k+i})$ as its input. Each LSTM cell consists of several densely-connected neural networks (NN). The actual output $h_t$ of the LSTM cell is a filtered subset of the cell state $S_t$. In the presented topology, the cell state $S_t$ and the output $h_t$ are arrays of the length $d$. The length $d$ defines the number of neurons in each neural network layer.

## 3　DATA ANALYSIS

### 3.1 Forward Model and Inverse Operator

MEG and MRI research data were collected after informed consent from a healthy 27-year old male under a protocol approved by the Massachusetts General Hospital Institutional Review Board. The subject had no medical history of hearing loss.

T1-weighted, high resolution MPRAGE (Magnetization Prepared Rapid Gradient Echo) structural images were acquired on a 1.5T Siemens whole-body MRI (magnetic resonance) scanner (Siemens Medical Systems) using a 32 channel head coil at MGH.

The structural data were preprocessed using FreeSurfer (Dale, Fischl, and Sereno 1999; Fischl, Sereno, and Dale 1999). After correcting for topological defects, cortical surfaces were tessellated using triangular meshes with ~130,000 vertices in each hemisphere. To expose the sulci in the visualization of cortical data, we used the inflated surfaces computed by FreeSurfer.

The dense triangulation of the folded cortical surface provided by FreeSurfer was decimated to a grid of 2,562 dipoles per hemisphere, corresponding to a spacing of approximately 6.2 mm between adjacent source locations. A piecewise-homogenous head conductor model with three compartments bounded by the inner skull, outer skull and outer skin was assumed, and the boundary element method (BEM) was used to compute the gain matrix (Hamalainen and Sarvas 1989). The watershed algorithm in FreeSurfer was used to generate the tessellations based on the MRI scan of the participant.

The initial current distribution estimate $\hat{q}_t$ was attained using the minimum-norm estimate (MNE) with loose current dipole orientation constraints set at 0.2, where 0.0 corresponds to fixed and 1.0 to free orientations. The regularized ($\lambda = 0.1$) noise covariance matrix used to calculate the inverse operator was calculated over the pre-stimulus period. To reduce the superficial bias of MNE, we incorporated depth weighting by adjusting the source covariance matrix, which has been shown to reduce point-spread (Lin et al. 2006), resulting in weighted MNE (wMNE). The dSPM estimate (Dale et al. 2000) was found by dividing wMNE with the projection of the estimated noise covariance matrix onto each source point. All forward and inverse calculations were done using MNE-C software (Hamalainen 2010; Gramfort et al. 2014).

### 3.2 Design of Simulation Study

A simulation study was done to test the performance of CMNE in comparison to dSPM, pure LSTM prediction and a control estimation. In the control estimation, 80 sequential dSPM distributions were averaged and multiplied with the dSPM of the following sample, thus mimicking the Markov Chain but without the LSTM network. The averaging can be seen as a simple predictor and since we use the same data as in our CMNE approach, the gain of using LSTM networks can be examined by comparing this control estimate with the CMNE. These simulations were designed to mimic the propagation of an epileptiform discharge in the left auditory cortex. The source configuration consisted of 5124 current dipoles placed over the cerebral cortex with free orientations. The activation time course of each of the three source dipole components over the cortical mantle were based on stationary Gaussian noise with spectral characteristics taken from EEG readings, as described in (Hunold et al. 2016). The activation time course of the five current dipoles in the superior temporal gyrus that model the epileptiform discharge were then superimposed by a spike-wave complex lasting for 200 ms, starting sequentially in the posterior-anterior direction and making one simulated epileptiform discharge lasting for a total duration of 1000 ms (Figure 3). The spike-wave complex had an amplitude 5 times larger than that of the background activity at 10 nAm. These five sources modeling epileptic foci were located 9 mm apart. A total of 250 epileptiform discharges were simulated with a 500 ms interictal period. A topographical view of the sensor level activity visualizing the average of all simulated spikes is shown in Figure 4.

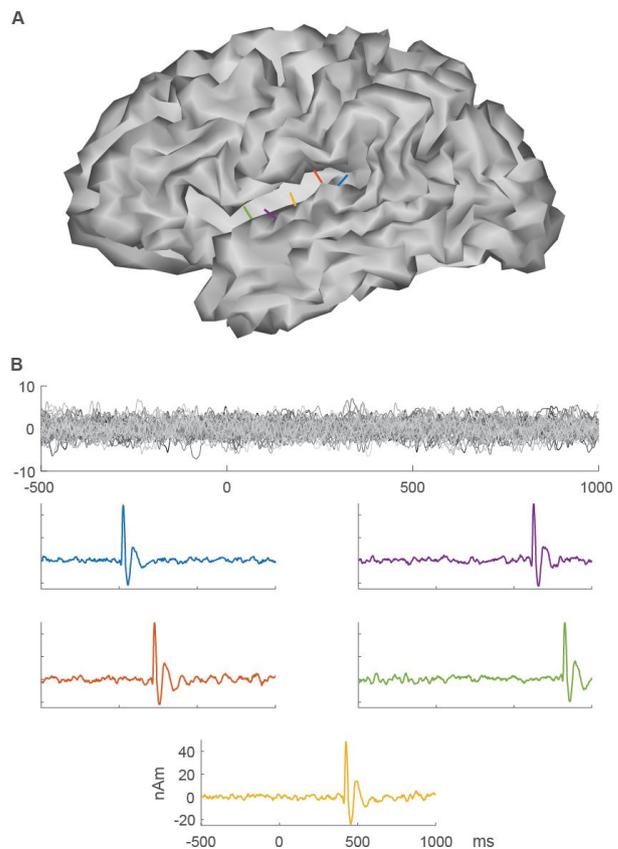

*Figure 3: Simulation setup. A) Gray matter surface with five current dipoles in the superior temporal gyrus modeling epileptic foci, color-coded according to their sequential activation pattern as shown in B. B) Time courses of source dipole amplitudes modeling epileptiform discharges (color-coded graphs) and background activity (upper graph). One epoch starts with an interictal period of only background activity lasting for 500 ms followed by a propagating epileptiform discharge lasting for 1000 ms.*

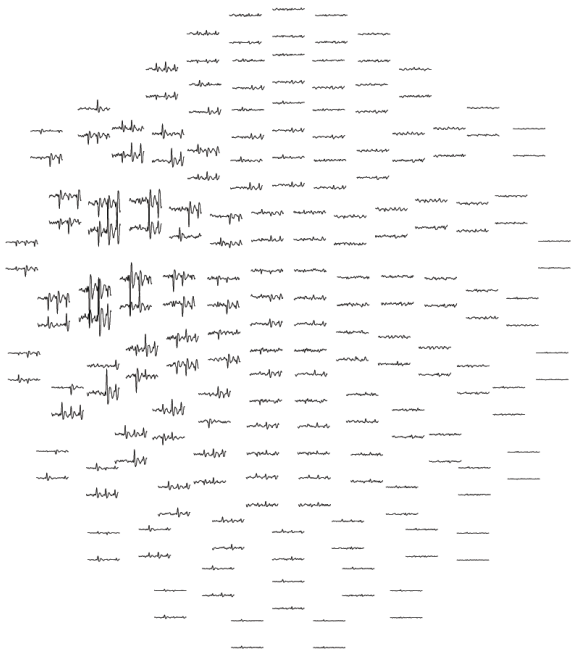

*Figure 4: Topographical view of simulated sensor level activity, visualizing averaged evoked responses using all simulated spikes.*

### 3.3 Auditory Steady State Response (ASSR) study

Auditory steady state response (ASSR) data were recorded from a subject in the MGH Martinos center MEG core in Charlestown, MA, using MEG and EEG. It is the same ASSR data that were used in (Samuelsson et al., n.d.). The MEG system was an Elekta Neuromag (Stockholm, Sweden) VectorView 306 channel MEG with 102 tri-pairs consisting of one magnetometer and two orthogonal planar gradiometers, making a total of 204 planar gradiometers and 102 magnetometers. The EEG was recorded with a 58 channel EasyCap system (EasyCap GmbH, Germany). The experiment was performed in a quiet, magnetically shielded room (Imedco, Switzerland). The sampling rate was 5000 S/s. Auditory steady state responses were elicited by an amplitude modulated (AM) sound with a duration of one second. The sound was played to the subject with an inter-trial pause of random and uniformly distributed 0.5-1.25 seconds duration. The carrier signal was a $f_0 = 1$ kHz sinusoid and was amplitude modulated to a depth of 90% by a superposition of a $f_1 = 40$ Hz and $f_2 = 223$ Hz sinusoid;

$$y(t) = (0.1 + 0.9[\sin(2\pi f_1 t) + \sin(2\pi f_2 t)]/2)\sin(2\pi f_0 t). \quad (9)$$

The choice of frequencies of the amplitude modulation signal was motivated by the fact that 40 Hz elicits a strong cortical response in the auditory cortex, located in the inferior wall of the Sylvian fissure (Hari, Hamalainen, and Joutsiniemi 1989). For frequencies higher than 150 Hz, the neural signal is dominated by upstream structures in the auditory pathway located in the brainstem (Bharadwaj and Shinn-Cunningham 2014). By having the two amplitude modulation signals superimposed, we elicit frequency tagged, simultaneous activity in one subcortical structure (223 Hz) and one in auditory cortex (40 Hz). The acquired data contained 1653 clean, i.e., artifact free, epochs.

### 3.4 Training and Evaluation

The LSTM networks were trained using sample-wise dSPM source estimates of single-trial epochs. The available epochs were divided up in one training data set (85%) and one validation data set (15%). The training- and validation data sets were thus disjoint. The training data were generated using overlapping sliding windows in time over the epochs, each window containing $k$ time steps – one for each LSTM cell. The LSTM network predicts the subsequent dSPM activation map based on these $k$ past time steps and the label is the actual dSPM estimation of the subsequent sample. Features are thus $k \times \rho$ dimensional, $\rho$ being the number of source points. Prior to the training, the features (past $k$ dSPM estimates) and labels (current dSPM estimate) were standardized by z-scoring, i.e., the mean was subtracted from the source estimates and they were divided by their respective standard deviation.

We employed the mean-square error (MSE) as the loss (objective) function and stochastic gradient descent (Adam) (Kingma and Ba 2014) as the optimization method. The training was organized in a minibatch setting which split the training process into small batches comprising a small set of gradient evaluations, the LSTM weights being updated using the anti-gradient of the error with respect to the LSTM weights over each minibatch.

The LSTM setup, training and evaluation was realized in CNTK (Seide and Agarwal 2016) / TensorFlow (Abadi et al. 2015) in combination with Keras (Chollet 2015) as the frontend API.

## 4    RESULTS

### 4.1 Network Architecture

The suitable hyperparameter set, i.e., the number of hidden units $d$ in the LSTM network and the number of past time steps $k$ used as inputs to the LSTM prediction, were determined empirically using the ASSR data set. We split the training data into 30 minibatches. Each minibatch consisted of 30 feature representations, each feature representation being a randomly selected window of 81 subsequent samples, 80 being used as inputs to the LSTM network to predict the 81st sample, which is compared to the label.

First, the influence of the LSTM units on the performance was tested by varying $d$ with a constant look back of $k = 80$ samples. The training results are shown in a loss graph in Figure 5. The loss is also known as the cost

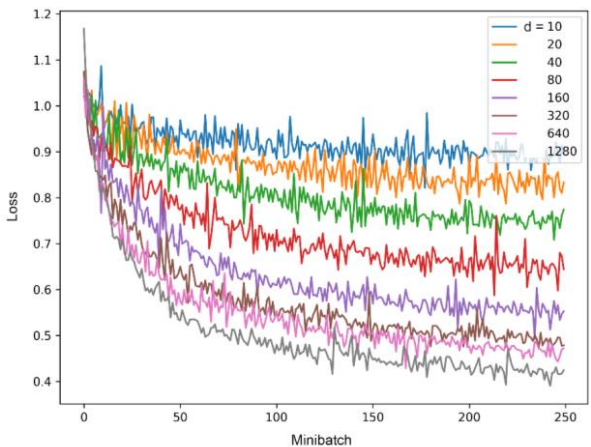

*Figure 5: Loss progression: influence of the number of hidden units $d$ in LSTM network on the performance with a constant look back of $k = 80$ samples.*

function where a lower loss indicates a higher prediction accuracy.

Second, the optimal number of past time steps $k$ was determined with a fixed number of hidden units $d = 640$, which is an appropriate trade-of between prediction

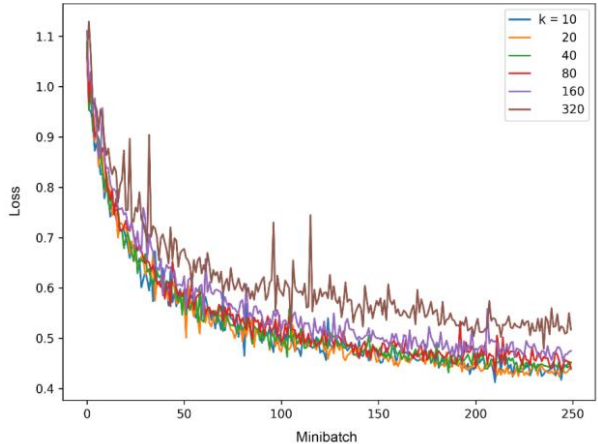

*Figure 6: Loss progression: influence of the look back, i.e., number of past samples $k$ used in the LSTM network with a fixed number of hidden units $d = 640$.*

accuracy and training time. The results are depicted in Figure 6.

Based on this empirical evaluation, we chose a final LSTM network topology of $d = 1280$ units and a window size comprising the past $k = 80$ samples. The selection of the number of hidden units was a compromise between training time and prediction accuracy. We anticipate that a larger number of units would further improve the network performance. The choice of $k$ was also a trade-off; shorter windows have fewer LSTM cells and thus fewer weights to adjust, leading to faster convergence, while wider time windows are more robust to fluctuations. $k = 80$ was found to be a good balance between the two.

## 4.2 Simulation Study

In the simulation study, we trained the LSTM network with the topology determined in Sec. 4.1 ($d = 1280, k = 80$) using only 100 minibatch iterations each comprising 30 evaluations containing 25 windows per evaluation. The windows contained 81 subsequent source estimation samples, whose starting points were randomly selected from the 212 raw training epochs.

Figure 7 shows a comparison between CMNE, dSPM, LSTM prediction and a control source estimation. It can be concluded that CMNE has the highest SNR in space and time as defined by the maximal current dipole amplitude in label A (second column, Figure 7). The source activation map shows that CMNE has the most focal source estimation, comparable to that of dSPM. Plotting the maximal current dipole amplitude of label A over time (third column, Figure 7) shows that the focal source estimate of CMNE remains over the entirety of the epoch. However, the LSTM prediction is not able to capture the temporal dynamics of the neural activation patterns adequately; rising slopes are blunted or left out when the dipole activation happens too fast. This inability of capturing fast activity changes gets more prominent the faster these changes are. The correction step that utilizes the current dSPM estimation compensates for this; the peak of the CMNE estimation ascent is on time but the ripples following the spike are suppressed. The control estimate has a reduced SNR, significantly lower than the SNR of CMNE, and distorts the amplitude relations.

Figure 8 shows source mapping of the simulated epileptiform discharge that was introduced in Figure 3. Since this activation pattern is in the form of dipoles at different locations activated subsequently, it should be highly predictable and the CMNE estimation should give better results over time as progressively more information on past activity is gathered, as compared to other estimation methods. This is indeed the case, as it can be seen from Figure 8 that while the dSPM mapping gets increasingly smeared out over time, the CMNE estimation remains focal with an adequate maximal dipole strength. The clear amplitude suppression of dipoles activated earlier also indicates spatial separation. The estimation error of the maximum activation compared to the ground truth is also improved.

This simulation outcome can be interpreted in analogy to a pre-surgical epilepsy study. In long-term monitoring, 200 ictal events have been observed, which can be categorized into five groups. From each group, 20 events have been averaged. These results show that CMNE anticipates the temporal propagation of the activity utilizing contextual information and separates the spatio-temporal distinct sources.

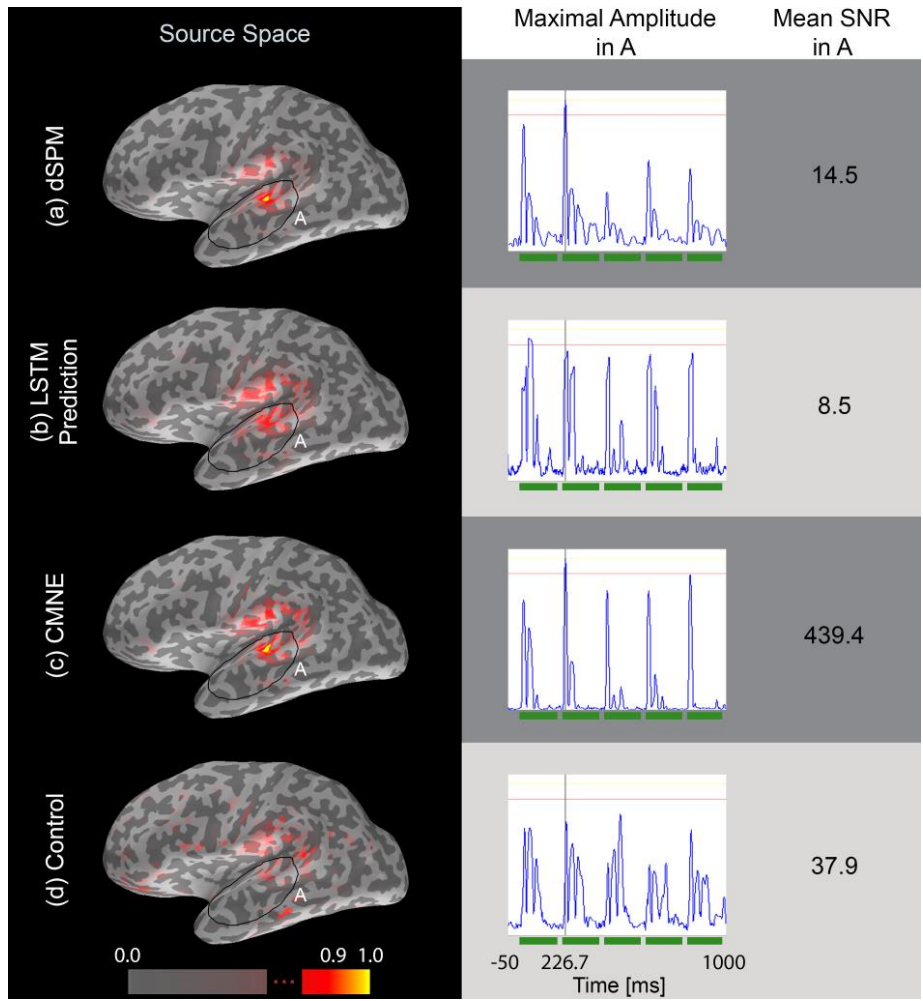

*Figure 7: Simulation results of 20 averaged epochs. Source estimations are shown in the source space column. The middle column shows the time traces of the dipole with maximal amplitude within the marked area A as depicted in the first column. The right column shows the estimated SNR of the respective source reconstruction method, defined as the maximal amplitude in label A during the sequences of activation (marked as green in the middle column) divided by the amplitude during the time in between activations. The tested source reconstruction methods were: a) dSPM, a noise-normalized MNE, b) prediction with an LSTM network based on $k = 80$ subsequent samples, c) Contextual MNE (CMNE) and d) the control estimation which averages the activity distributions of 80 previous subsequent samples and multiplies them with the activity distribution of the current sample.*

### 4.3 Auditory Steady State Response Study

CMNE was evaluated based on empirical ASSR data in the same way as was done in the simulation study; the same LSTM network topology was used ($k = 80, d = 1280$) and the results were compared with dSPM, the LSTM prediction alone and the control estimate. The training was performed with 250 minibatch iterations each comprising 30 evaluations containing 20 windows per evaluation. The windows contained 81 subsequent source estimation samples, which were randomly selected from the 1405 artifact-free raw training epochs.

Validation was made based on the remaining 248 epochs that were not used in the training. Figure 9 shows a comparison between source mapping of the validation ASSR data done by CMNE, dSPM, LSTM prediction alone and a control estimation, and are based on 20 averaged epochs. The label A1 marks the primary auditory cortex.

Consistent with the simulation study, CMNE gives the most focal source estimate. CMNE also shows an improved SNR over time, which makes sense given the contextual nature of the LSTM networks, resulting in the SNR being one order of magnitude higher than that given by dSPM. The control estimation has the lowest SNR.

## 5   DISCUSSION

Intracranial electrophysiology studies have shown that brain activity is spatially and temporally correlated (Bullock et al. 1995; Destexhe, Contreras, and Steriade 1999; Leopold, Murayama, and Logothetis 2003; Nunez 1995). Studies using fMRI have also shown that brain activity is temporally coherent in spatially distributed networks (Fox and Raichle 2007; Fox et al. 2005; Gusnard and Raichle 2001; Raichle et al. 2001). These findings inspired the development of CMNE, which introduces temporal priors and thus takes the temporal context of brain activity into account. This approach adds additional information to the inverse problem and thereby reduces its ill-posedness.

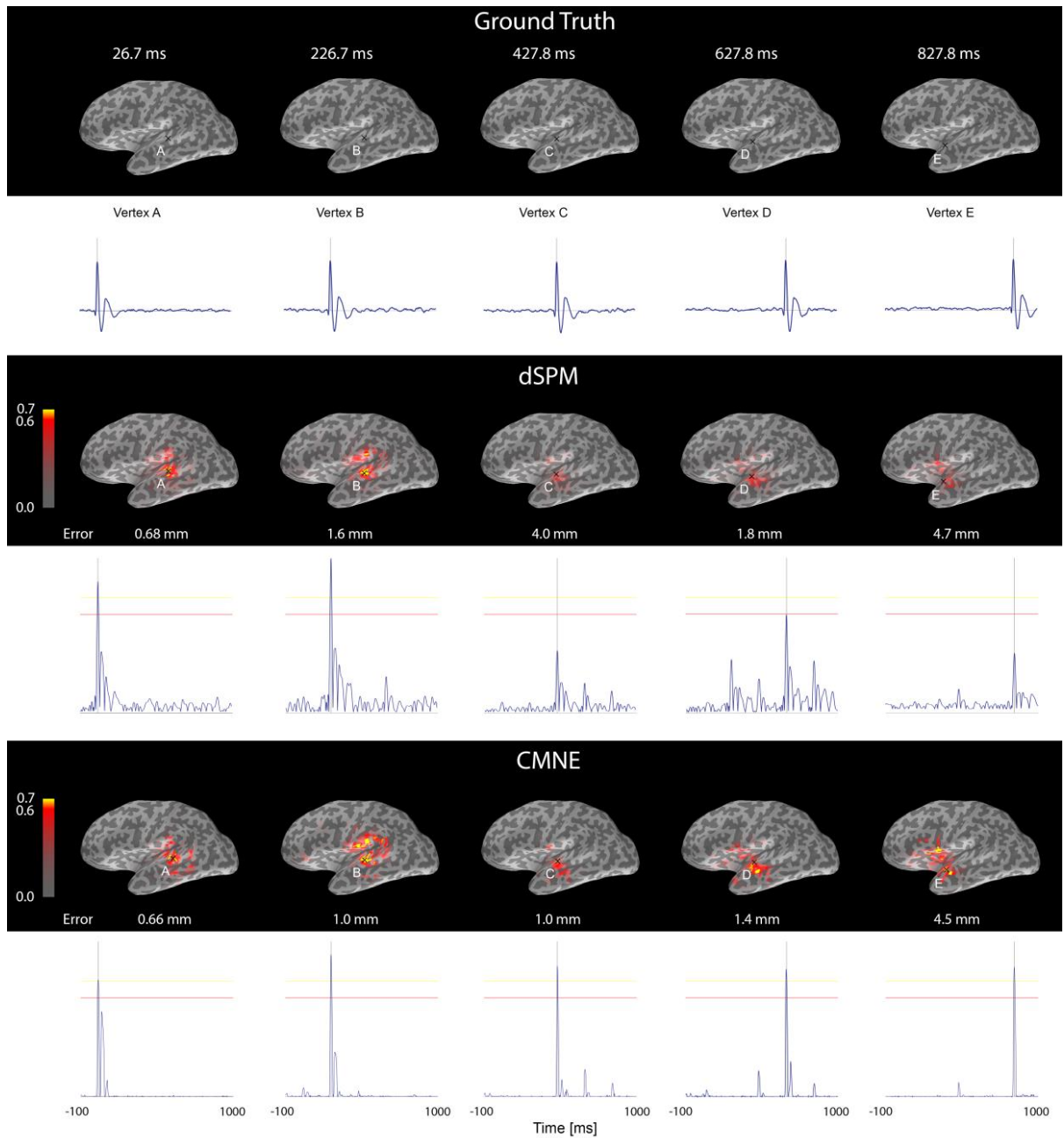

*Figure 8: Source mapping of simulated epileptiform discharge data. dSPM (middle row) and CMNE (lower row) source reconstructions displayed over an inflated brain surface in comparison to the ground truth (upper row) for different time points along with the time trace of the dipole with the greatest amplitude at that point in time (marked A-E). The estimation error indicates the distance between the activated dipole location and the location of the dipole with the greatest amplitude. The results are based on an average of 20 epochs.*

CMNE is a novel distributed source estimation method belonging to the family of methods based on artificial neural networks, see e.g., (Jun and Pearlmutter 2005; Latvala 2017). The present approach also relates closely to the work of (Lamus et al. 2012) which was based on Kalman filters, a special case of Markov localization. The work of (Lamus et al. 2012) can model locally distributed immanent connections between two samples, since activation is propagated with a predictive kernel $\mathcal{F}$: $\hat{q}_t = \mathcal{F}\,\hat{q}_{t-1}$. This propagation implicitly includes the whole estimation history similarly to CMNE. A limitation of the Kalman approach (Lamus et al. 2012) is that past activity is directly manifested in the subsequent source estimation.

CMNE with its LSTM networks avoid this issue since LSTMs distinguish between an internal state $S_t$ and the current output state $h_t$. This allows for the preservation of information over a long range of time without being manifested in subsequent samples. Storing information in a hidden state enables embedding source activity distributions in the context of neural processing, e.g., widely connected networks form activation patterns which are not always detectable in, e.g., MEG measurements. The Markov localization is arranged in a chain and incorporates all past information forming an infinite chain which extends beyond the last $k$ samples.

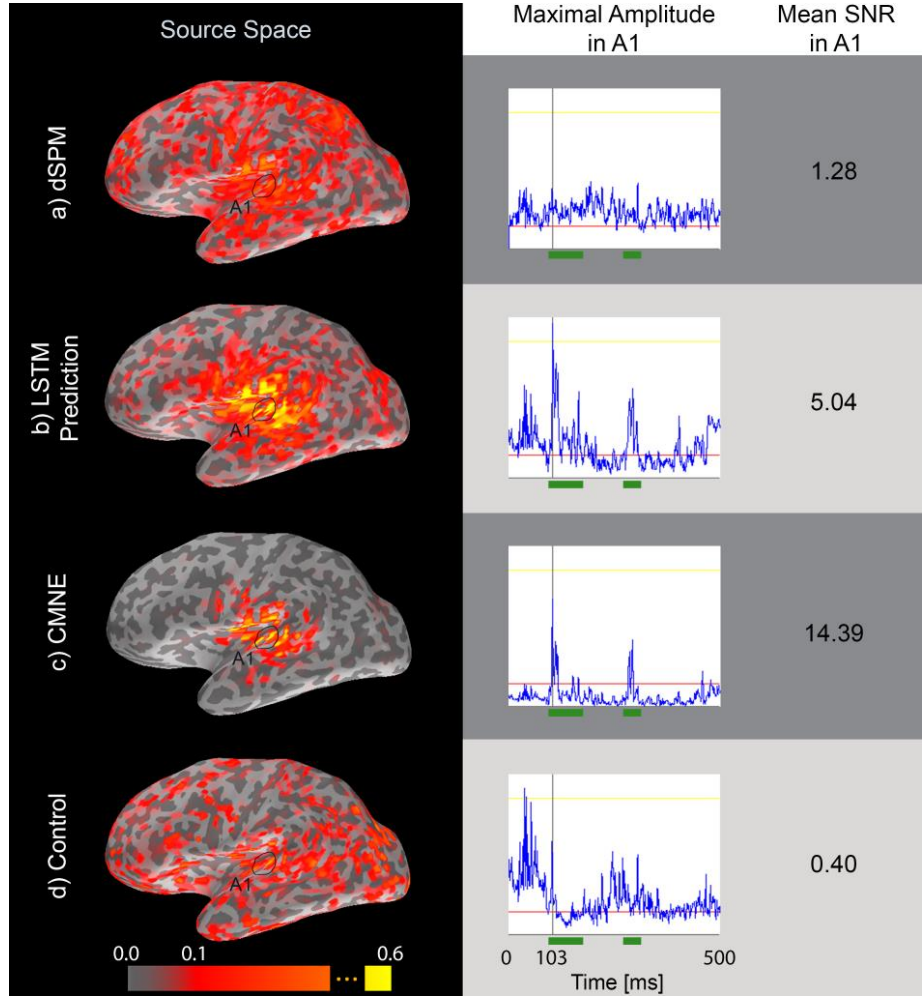

*Figure 9: Mapping of ASSR data based on averages of 20 epochs. Source estimates (left column), time courses (middle column) of the source dipole in A1 with the greatest amplitude along with SNR (right column). SNR was estimated by the time ranges marked as green in the middle column as signal, and the rest as noise. Each row represents one source estimation method: (a) dSPM, (b) LSTM prediction based on 80 samples, (c) Contextual MNE, and (d) control estimation.*

As anticipated, simulation results showed that CMNE estimations improve when based on a longer time horizon. However, the LSTM predictions themselves were not always able to follow steep ascends in the simulated data, which relates to a lowpass characteristic. The later wave forms in the simulated data were most likely not present in the LSTM prediction because of this limitation. The correction step with dSPM results in a better representation of the temporal characteristics and even increased SNR.

These promising results obtained with CMNE lead us to believe that the hidden LSTM cell state $S_t$ can be useful in different ways. Based on the high prediction accuracy, we suspect that the LSTM networks abstract the functional connectivity for the elicited neural response that the network is trained on. In fact, the LSTM cell state $S_t$ could in this case be interpreted as a state abstracting the underlying brain activity, and CMNE could be used to relate or integrate different imaging modalities by combining the internal state $S_t$ of different transition models. This approach would be comparable to the work of (Cichy, Pantazis, and Oliva 2014; Cichy et al. 2016, 2017; Ravi et al. 2017), who related different modalities by means of representational similarity matrices.

## 6  CONCLUSION

CMNE incorporates contextual information to help solve the inverse problem in MEG by making it less ill-posed. CMNE uses LSTM networks to predict source activity based on the preceding activation patterns. These predicted source activations are then corrected by a dSPM estimation based on the current measurement. Using simulations of ictal events and real ASSR data from a human subject, it was shown that CMNE can achieve high prediction accuracy and could thus be a valuable addition to the family of inverse methods in M/EEG.